# Universal behavior of the upper critical field in iron based superconductors

J. L. Zhang, L. Jiao, Y. Chen, H. Q. Yuan

Department of Physics, Zhejiang University, Hangzhou, Zhejiang 310027, China

**Abstract**

The newly discovered iron-based high temperature superconductors have demonstrated rich physical properties. Here we give a brief review on the recent studies of the upper critical field and its anisotropy in a few typical series of the iron-base superconductors (FeSCs). In spite of their characters of a layered crystal structure, all the FeSCs possess an extremely large upper critical field and a weak anisotropy of superconductivity, being unique among the layered superconductors. These particular properties indicate potential applications of the FeSCs in the future. Based on the experimental facts of the FeSCs, we will discuss the possible mechanisms of pair breaking in high magnetic fields and its restrictions on the theoretical analysis of the superconducting pairing mechanisms.

# 1. Introduction

The recent discovery of the iron-based superconductors with $T_c$ as high as 55K has attracted world-wide interests [1-6]. Partially benefited from the previous experience gained in the study of the high-$T_c$ cuprates, various families of FeSCs have been subsequently synthesized and investigated in the past three years. Typical series of the FeSCs include ReOFeAs(O,F) ( Re=La, Ce, Pr, Nd, Sm or Gd, 1111-type) [1-6], $AFe_2As_2$ (A=alkaline or alkaline-earth metals, 122-type) [7,8], α-PbO type FeSe/FeTe (11-type) [9,10], AFeAs (A=Li, Na, 111-type) [11-13], and $AFe_xSe_2$ (A=K, Cs, Rb, $(Tl_{1-y}K_y)$ and $(Tl_{1-y}Rb_y)$) [14-18]. All these compounds share a common feature in their crystal structures, i.e., it is stacked with repeated Fe-As or Fe-Se(Te) layers. Such a layered crystal structure and the high superconducting transition temperature of the iron pnictides/chalcogenides bear many similarities to the high-$T_c$ cuprates. Therefore, it is nature to compare the physical properties of the FeSCs with those of high-$T_c$ cuprates, which might provide an alternative to study the mysteries of high-$T_c$ superconductivity.

However, a growing number of evidence has demonstrated that significant difference may exist between the FeSCs and the high-$T_c$ cuprates. According to our knowledge, the major discrepancies can be summarized as follows: (i) The parent compounds of the FeSCs are usually a bad metal instead of an antiferromagnetic Mott-insulator as observed in the cuprates [19]. Note that the $AFe_xSe_2$ superconductors are likely associated with a Mott-insulate parent compound [14-18], but its nature is still under debate; (ii) A d-wave pairing state was realized in the high-$T_c$ cuprates, but an S±-type (or S++ type) order parameter has been proposed for the FeSCs [20, 21]; (iii)The FeSCs show multi-band electronic structures, while the cuprates are a single-band system [19]; (iv) The FeSCs show a weak anisotropy of superconductivity in spite of their layered crystal structures.

The upper critical field, $\mu_0H_{c2}$, is one of the fundamental parameters in type II superconductors, which provides important insights on the pair-breaking mechanisms in a magnetic field. Furthermore, other superconducting parameters, e.g., the coherence length and the anisotropic parameter, can be derived from the upper critical field. It has been found that the iron-based superconductors usually possess an extremely large upper critical field and the traditional extrapolation of $\mu_0 H_{c2}(T_c)$ near $T_c$ to low temperatures usually gives a wrong estimation. Therefore, a large magnetic field, which usually requires the facilities of pulsed magnetic fields, is desired for studying the upper critical field of the FeSCs.

This article will provide a brief overview on the universal behavior of the upper critical field and its anisotropy in FeSCs. Following the introduction, we will describe the crystal structures of various families of the FeSCs and the basic mechanisms for superconducting-pair breaking in a magnetic field. Then, we will devote to the upper critical field and its anisotropy of several typical FeSCs. The article will end with a summary and prospects after comparing the upper critical field of the FeSCs with other layered superconductors.

## 2. Layered crystal structures of iron-based superconductors

Resembling the high-$T_c$ cuprates, the iron based superconductors are layered compounds which crystal structures are stacked with FeAs or FeSe layers. Various atoms or molecules can be intercalated between the FeAs or FeSe layers to modulate the lattice constants or even to change the crystal structures, leading to the formation of various FeSCs [1-18]. On the other hand, the intercalated atoms or molecules may also act as the donators of charge carriers, giving rise to the hole- or electron-doped superconductors [1-18]. As an example, we show the crystal structures and the lattice parameters of the major types of FeSCs in Fig. 1 and Table 1. One can see that all the compounds possess the same FeAs or FeSe layers in their crystal structure. Different from the high-$T_c$ cuprates in which Cu and O atoms are located on a square planar, in FeSCs the Fe and As (or Se) atoms form $FeAs_4$ (or $FeSe_4$) tetrahedrons. It has been shown that the superconducting transition temperature $T_c$ of FeSCs depends on the angles between the FeAs (or FeSe) bonds [22] and the height of the tetrahedrons [23].

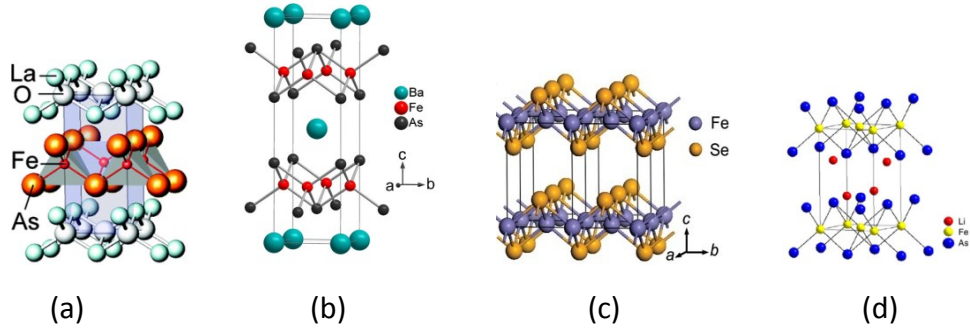

(a) (b) (c) (d)

Fig.1. The crystal structures of several typical FeSCs: (a) LaOFeAs [1], (b) BaFe$_2$As$_2$ [7], (c) α-FeSe [9] and (d) LiFeAs [12].

The 1111-families of FeSCs crystallize in a tetragonal structure with a P4/mmn space group [1-6], in which the FeAs and LaO layers are alternately arranged. A record of $T_c$ (≈ 55K) of the FeSCs was achieved in the 1111-series [5,6]. BaFe$_2$As$_2$, an oxygen free compound, belongs to I4/mmm space group and crystallizes in the ThCr$_2$Si$_2$–structure [7]. Note that in each unit cell there contains two FeAs layers in BaFe$_2$As$_2$, but only one FeAs layer in LaOFeAs. The Fe(Se,Te) compounds are formed by a stack of edge sharing FeSe$_4$ (FeTe$_4$) layers without a charge reservoir which makes it unique among the FeSCs [9, 10]. In all the above FeSCs, their parent compounds undergo a structural transition and antiferromagnetic transition upon cooling down from room temperature, showing a metallic ground state [24-26]. Superconductivity develops while suppressing the antiferromagnetic/structural phase transition order by elemental substitutions [1-8] or by applying pressure [27, 28]. On the other hand, the stoichiometric compounds LiFeAs and NaFeAs become superconducting at ambient pressure [11-13]. In these compounds, Li (or Na) is sandwiched between the FeAs layers. Very recently, superconductivity was also observed in the iron selenides AFe$_x$Se$_2$ (A=K, Rb, Cs, Tl) with $T_c$ up to 33K [14-18]. These compounds crystallize in the same ThCr$_2$Si$_2$–structure as those of AFe$_2$As$_2$, but the Fe sites are only partially occupied in order to maintain the electric neutrality, leading to the formation of iron vacancy order. Being different from other FeSCs, superconductivity in AFe$_x$Se$_2$ seems to be associated with an antiferromagnetic Mott insulator [29-32] and the vacancy tuned superconductivity has been proposed as a new approach for obtaining high-$T_c$ superconductors [33]. However, it remains highly controversial on whether magnetism and superconductivity coexist[30, 31] or undergo a phase separation in the iron selenides[34, 35].

Table 1: The crystal structures and lattice parameters of several typical FeSCs.

| Family | Compounds | Optimal $T_c$ (K) | Lattice parameters (nm) | c/a | space group at RT | Ref. |
|---|---|---|---|---|---|---|
| 1111 | LaOFeAs | 26K | a=0.403552(8)<br>c=0.87393(2) | 2.175 | P4/*nmm* | [1] |
| | SmOFeAs | 55K | a=0.3933(5)<br>c=0.8495(4) | 2.137 | P 4/nmm | [5] |
| 122 | BaFe$_2$As$_2$ | 38K | a= 0.39625(1)<br>c= 1.30168(3) | 3.285 | I4/*mmm* | [7] |
| | K$_x$Fe$_x$Se$_2$ | 30K | a=0. 39136(1)<br>c=1.40367(7) | 3.59 | I4/mmm | [13] |
| | (Tl,Rb)Fe$_x$Se$_2$ | 32K | a=0.3896<br>c= 1.4303 | 3.61 | I4/mmm | [18] |
| 111 | LiFeAs | 18K | a=0.37914(7)<br>c=0.6364(2) | 1.678 | P4/nmm | [11] |
| 11 | Fe(Se,Te) | 8-13K | a=0.3765<br>c=0.5518 | 1.466 | P4/nmm | [9] |

In general, all the Fe based superconductors share a common feature of possessing the FeAs or FeSe layers in their crystal structures. The conducting carries are restricted to the FeAs or FeSe layers, in which the Fe-ions are tetrahedrally coordinated with pnictogen (P, As) or chalcogen (Se, Te) ions and form a square planar. As a result of the layered crystal structure, the electronic structure is expected to be anisotropic, as seen in some band structure calculations [36, 37] and the anisotropic resistivity in the normal state [38]. However, these compounds display remarkable isotropic superconducting properties.

### 3. Magnetic field induced pair breakings

In a superconductor, two electrons can be attracted together to form a bound state, i.e., the Cooper pair, in a certain manner. Application of an external magnetic field may destroy the Cooper pairs in the following two ways: i) the orbital pair breaking due to the Lorentz force acting via the charge on the paired electrons, known as the orbital limit; (ii) the Pauli paramagnetic pair breaking as a result of the Zeeman effect which aligns the spins of two electrons with the applied field, called as the Pauli paramagnetic limit.

Werthamer, Helfand and Honenberg (WHH) systematically studied the temperature and impurity dependence of the upper critical field for type II superconductors in 1960s [39], in which the effects of both Pauli paramagnetism and spin-orbit scatterings were considered. These factors largely complicate the general expression of $\mu_0 H_{c2}(T_c)$ and usually one can only consider some simplified cases while fitting the experimental results.

In the dirty limit, the upper critical field $\mu_0 H_{c2}$ can be expressed in terms of the digamma function [39]:

$$\ln\frac{1}{t} = \left(\frac{1}{2} + \frac{i\lambda_{so}}{4\gamma}\right)\psi\left(\frac{1}{2} + \frac{\hbar + \frac{1}{2}\lambda_{so} + i\gamma}{2t}\right) + \left(\frac{1}{2} - \frac{i\lambda_{so}}{4\gamma}\right)\psi\left(\frac{1}{2} + \frac{\hbar + \frac{1}{2}\lambda_{so} - i\gamma}{2t}\right) - \psi\left(\frac{1}{2}\right), \quad (1)$$

where

$$t = T/T_c, \quad \gamma \equiv \left[(\alpha\hbar)^2 - \left(\frac{1}{2}\lambda_{SO}\right)^2\right]^{\frac{1}{2}}, \quad (2)$$

and

$$\hbar = -\frac{4\mu_0 H_{c2}}{\pi^2 (dH_{c2}/dt)|_{t=1}}. \quad (3)$$

Here the parameter $\lambda_{so}$ describes the strength of the spin-orbit scattering and $\alpha$ is so-called Maki parameter (see below). One can numerically solve Eq. (1) to obtain the temperature dependence of the upper critical field. Practically, we can analyze the experimental data of $\mu_0 H_{c2}(T_c)$ in terms of the WHH theory by adjusting the fitting parameters of $\lambda_{so}$ and $\alpha$.

In the absence of spin paramagnetic effect ($\alpha=0$), the upper critical field is then restricted by orbital pair breaking effect. In the weak-coupling case, Eq. (1) can be simplified as (assuming $\lambda_{so}=0$):

$$\ln\frac{1}{t} = \psi\left(\frac{1}{2} + \frac{\hbar}{2t}\right) - \psi\left(\frac{1}{2}\right). \quad (4)$$

The orbital limit of the upper critical field can then be derived as:

$$\mu_0 H_{c2}^{orb}(0) = -0.69 T_c \left(\frac{d\mu_0 H_{c2}}{dT}\right)|_{T=T_c}. \quad (5)$$

In the extreme Pauli limiting case, superconductivity is suppressed in a magnetic field when the spin polarization energy exceeds the superconducting condensation energy [40-42]. The Pauli limiting field in a weakly coupled superconductor, also called as Chandrasekhar-Clogston limit, is determined by the superconducting energy gap [40, 41]：

$$\mu_0 H^p(0)[T] = \Delta/\sqrt{2}\mu_B \qquad (6)$$

$$= 1.86 T_c[K] \text{ (for BCS SC)}. \qquad (7)$$

This limit can be enhanced in the cases of strong electron-phonon coupling, spin-orbital coupling or spin-triplet pairing state.

When the orbital limit of $\mu_0 H_{c2}^{orb}(0)$ and the Pauli limit of $\mu_0 H^p(0)$ are comparable, the paramagnetically limited effect may become crucial on determining the actual upper critical field, which can be expressed as [40-42]:

$$\mu_0 H_{c2}^P(0) = \mu_0 H_{c2}^{orb}(0)/\sqrt{1+\alpha^2}, \qquad (8)$$

where the Maki parameter $\alpha$ is defined as [43]:

$$\alpha = \sqrt{2} H_{c2}^{orb}(0)/H^p(0). \qquad (9)$$

The Maki parameter $\alpha$ represents the relative strength of orbital and spin pair breaking. A spatially non-uniform superconducting state, i.e., the Fulde-Ferrell-Larkin-Ovchinnikov (FFLO) state [44,45], may develop near the upper critical field provided that it possesses a sufficiently large $\alpha$ and is in the clean limit. In conventional superconductors, $\mu_0 H^P(0K)$ is usually larger than $\mu_0 H_{c2}^{orb}(0K)$ and, therefore, their upper critical field is mainly restricted by the orbital pair-breaking mechanism. However, spin paramagnetic effect may become dominant for pair breaking in unconventional superconductors, e.g., the heavy fermion superconductors and the organic superconductors.

**4. The upper critical field in iron based superconductors**

Soon after the discovery of superconductivity in iron pnictides, it was noticed that the FeSCs possess an extremely large upper critical field. Estimations of $\mu_0 H_{c2}(0)$ from its initial slopes near $T_c$ using the WHH model give a value as high as 100T~300T [46-48]. As an example, Fig.2 shows the upper critical fields $\mu_0H_{c2}(T_c)$ of the single crystalline NdFeAsO$_{0.82}$F$_{0.18}$ ($T_c$=49K) [48] and Ba$_{0.55}$K$_{0.45}$Fe$_2$As$_2$ ($T_c$=30K) [49] near $T_c$ which were determined from the resistive measurements. In both compounds, the upper critical field $\mu_0H_{c2}(T_c)$ shows a weak upward curvature with a large initial slopes. Extrapolation of $\mu_0H_{c2}(T_c)$ near $T_c$ to zero temperature using the WHH model gives $\mu_0H_{c2}^{H\|ab}(0K) = 304T$ (110T) and $\mu_0H_{c2}^{H\|c}(0K) = 70T$ (75T) for NdFeAsO$_{0.82}$F$_{0.18}$ (Ba$_{0.55}$K$_{0.45}$Fe$_2$As$_2$), respectively. However, these values are overestimated as shown below. To experimentally confirm them, it is highly desired to precisely determine the upper critical field down to much lower temperature, for which a pulsed magnetic field is required.

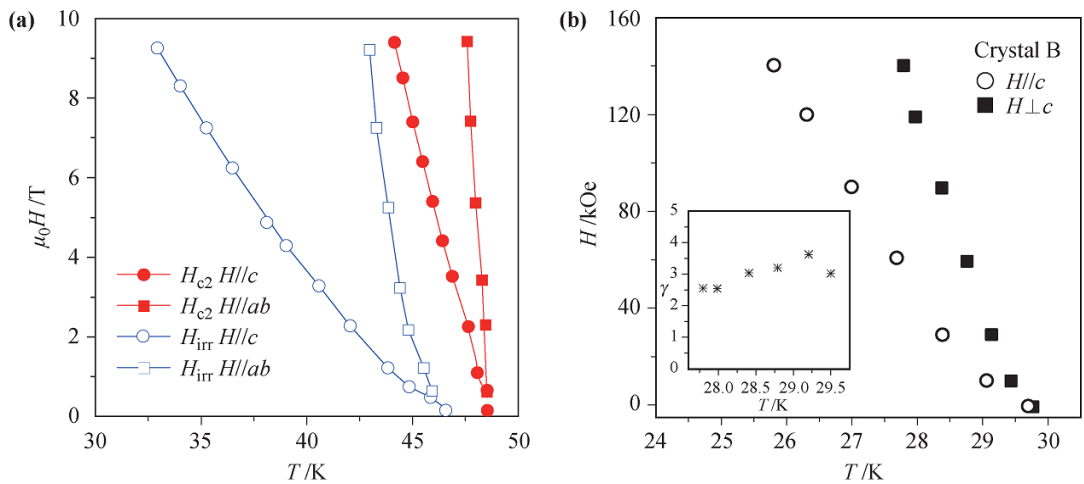

Fig.2 The upper critical field $\mu_0H_{c2}(T)$ at temperatures near $T_c$: (a) NdFeAsO$_{0.82}$F$_{0.18}$ [48] and (b) Ba$_{0.55}$K$_{0.45}$Fe$_2$As$_2$ [49].

We were one of the groups who first studied the upper critical field of FeSCs using a pulsed magnetic field immediately after its discovery. In Fig. 3, we show the field dependence of the frequency shift for LaFeAsO$_{0.9}$F$_{0.1-\delta}$, which was measured by a tunnel-diode oscillator (TDO) technique [47]. The derived upper critical fields

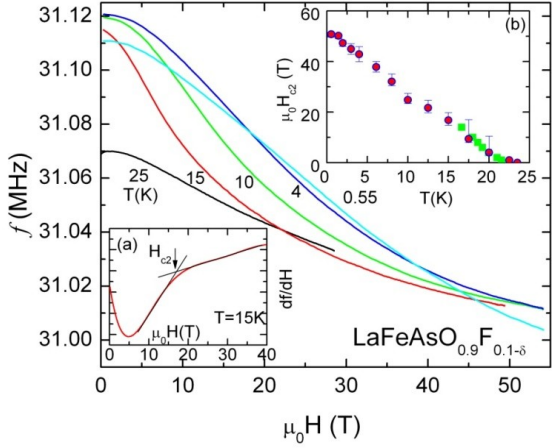 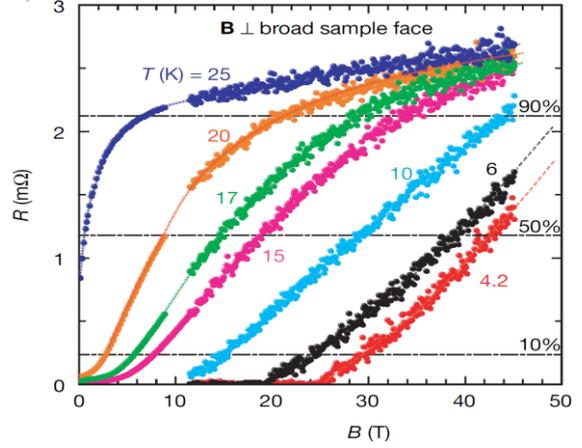

Fig.3 Field dependence of the TDO frequencies at various temperatures (from Ref [47]). Inset (a) shows the derivative of the TDO frequency with respect to field. Inset (b) plots the derived $\mu_0H_{c2}(T_c)$.

Fig.4 Magnetic field dependence of the electrical resistivity at various temperatures for LaFeAsO$_{0.89}$F$_{0.11}$ [50].

$\mu_0H_{c2}(T_c)$ are plotted in the inset (b) of Fig. 3, showing an almost linear temperature dependence with a zero temperature value of $\mu_0H_{c2}(0)\approx50$T. Simultaneously, Hunte et al measured the electrical resistivity of the polycrystalline LaFeAsO$_{0.89}$F$_{0.11}$ up to 45T [50], which shows a very broad superconducting transition (see Fig. 3). By assuming that such a broad resistive transition is attributed to the anisotropy of its upper critical field, in a polycrystalline sample Hunte et al derived two critical lines from the onset and the end point of the superconducting transitions, i.e., $B_{max}(T_c)$ and $B_{min}(T_c)$ (see Fig. 5(a)), which presumably correspond to the upper critical fields for field perpendicular and parallel to the c-axis, respectively[43]. $B_{min}(T_c)$ increases linearly with decreasing temperature near $T_c$ and then shows a significant upward curvature. This feature resembles that of MgB$_2$ and, therefore, was suggested as the first experimental evidence of two band superconductivity for the FeSCs [50]. Such an assumption of two-band superconductivity was later confirmed by ARPES experiments[51-53] and the anisotropic behavior of the upper critical field seems to be compatible with the subsequent studies on the single crystalline samples of the 1111-series. In Fig. 5, we plot the upper critical field $\mu_0H_{c2}(T_c)$ for several 1111-compounds: (a) polycrystalline LaFeAsO$_{0.89}$F$_{0.11}$ as described above[50]; (b) single crystal NdFeAsO$_{0.82}$F$_{0.18}$ [54] and (c) single crystal SmFeAsO$_{0.85}$ [55]. At a first glance, $\mu_0H_{c2}(T_c)$ of these compounds behave quite distinctively, in particular for that of SmFeAsO$_{0.85}$. Such a discrepancy might originate from the different doping contents and the multi-band electronic structure, but its origination remains unclear. Nevertheless, both the polycrystalline LaFeAsO$_{0.89}$F$_{0.11}$ and the single crystal NdFeAsO$_{0.82}$F$_{0.18}$ show a similar upward curvature for H//c. Furthermore, all these compounds show a very large upper critical field $\mu_0H_{c2}(0)$ and the curves for H//c and H⊥c tend to get closer at low temperature, indicating a decrease of the superconducting anisotropy with decreasing temperature.

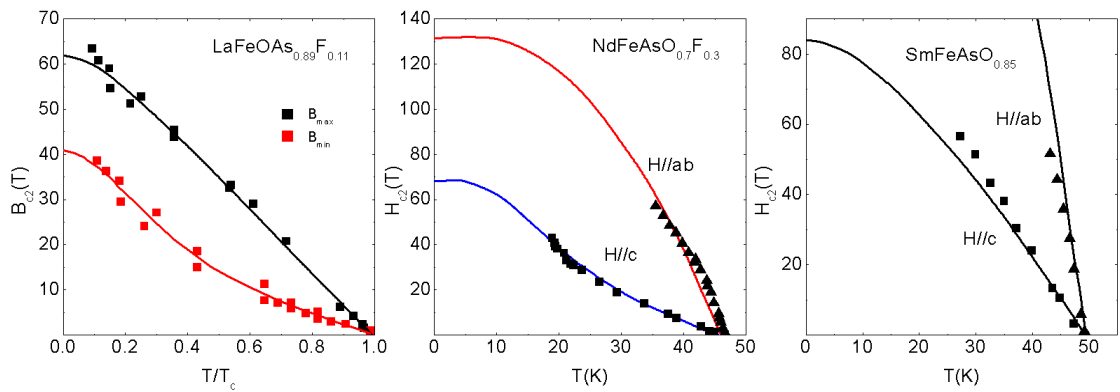

Fig.5 Temperature dependence of the upper critical field for polycrystalline LaFeAsO$_{0.89}$F$_{0.11}$ [50], single crystalline NdFeAsO$_{0.82}$F$_{0.18}$ [54] and SmFeAsO$_{0.85}$ [55].

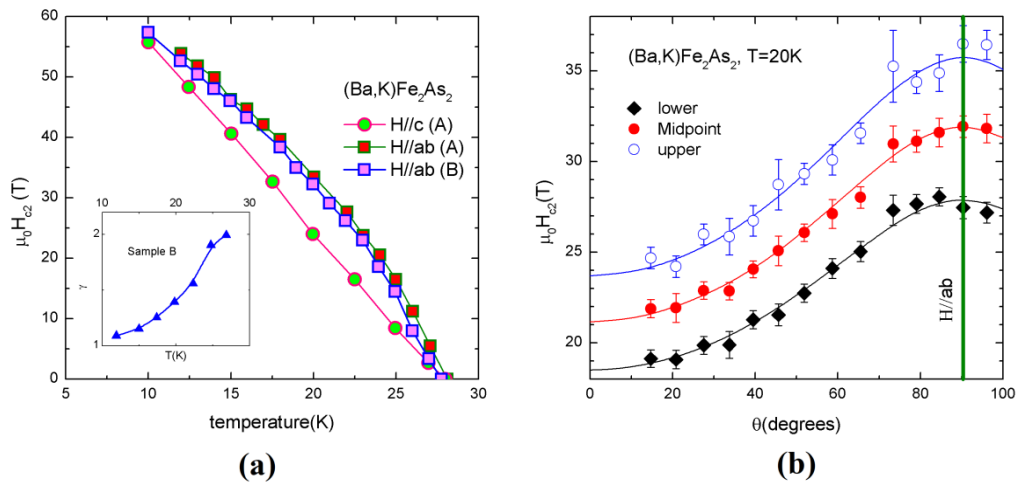

Fig.6 (a): The H-T phase diagram for single crystal (Ba$_{0.6}$K$_{0.4}$)Fe$_2$As$_2$. The inset shows the anisotropic parameter γ as function of temperature. (b): The angle dependence of the upper critical field $\mu H_{c2}(\theta)$ at 20K (from Ref [56]).

Nearly isotropic superconductivity was first realized in the hole doped compound (Ba$_{1-x}$K$_x$)Fe$_2$As$_2$ [56]. In Fig. 6, we plot the upper critical field μ$_0$H$_{c2}$(T$_c$) for the single crystalline (Ba$_{1-x}$K$_x$)Fe$_2$As$_2$. A prominent feature is that the upper critical field μ$_0$H$_{c2}$(T$_c$) increases linearly with decreasing temperature for H//c, but shows a concave curvature for H//ab. The two curves of $\mu_0 H_{c2}^{H\|c}(T_c)$ and $\mu_0 H_{c2}^{H\|ab}(T_c)$ eventually emerge together at low temperatures, suggesting an isotropic upper critical field in (Ba$_{1-x}$K$_x$)Fe$_2$As$_2$ which is striking for a layered superconductor. In order to better characterize the anisotropy of the upper critical field, we show the angle dependence of μ$_0$H$_{c2}$(θ) at 20K (see Fig.6 (b)), which can be well scaled by the single band anisotropic Ginzburg-Landau (G-L) theory [57]:

$$\mu_0 H_{c2}^{GL}(\theta) = \mu_0 H_{c2}^{H\|c} / \sqrt{\cos^2(\theta) + \gamma^{-2}\sin^2(\theta)}. \tag{10}$$

Here θ is the angle between the magnetic field and the c-axis. The anisotropic parameter γ is defined by:

$$\gamma = \sqrt{\frac{m_{ab}}{m_c}} = H_{c2}^{H\perp c} / H_{c2}^{H\|c}, \tag{11}$$

where m$_{ab}$ and m$_c$ are the effective masses of electrons for the in-plane and out-of-plane motion, respectively. It has been shown that a single band anisotropic model can properly describe the angular dependence of μ$_0$H$_{c2}$(θ) in a multi-band system at temperatures near T$_c$ [58]. From Fig. 6 (b), one can see that the experimental data of μ$_0$H$_{c2}$(θ) can be nicely fitted by Eq. 10 (solid lines), indicating that the anisotropic upper critical field is attributed to the effective mass anisotropy. The fittings give an anisotropic parameter of γ=1.5 which is in consistence with that directly calculated from $\gamma = H_{c2}^{H\perp c} / H_{c2}^{H\|c}$. Such a weak anisotropy of the upper critical field in (Ba$_{1-x}$K$_x$)Fe$_2$As$_2$ is in sharp contrast to other layered superconductors. In spite of the quasi-two dimensional crystal structure, its superconducting coherence length (ξ$_{ab}$(0)=ξ$_c$(0)=2.17nm, according to the upper critical field derived in Ref [56]) is much longer than the distance between the neighbor FeAs layers (0.32 nm) [7], indicating the importance of the interlayer interactions. In the following, we will show that such a weak anisotropy of $\mu_0H_{c2}(T_c)$ is a universal behavior of the FeSCs (see Fig. 7).

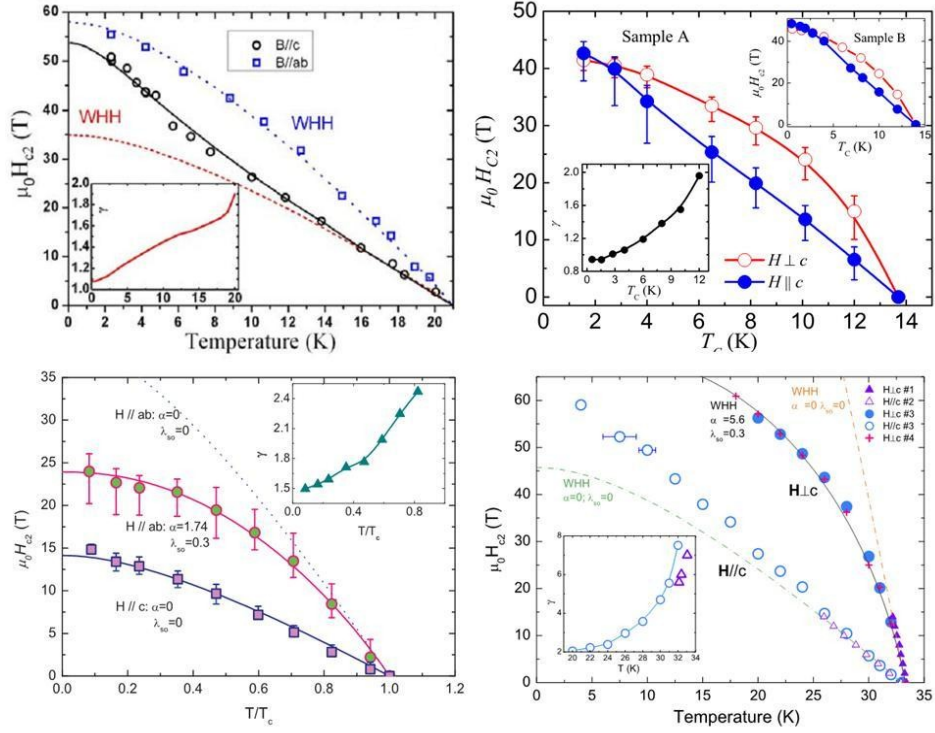

Fig.7 Temperature dependence of the upper critical field for (a) Ba(Fe$_{0.93}$Co$_{0.07}$)$_2$As$_2$ [59], (b)Fe$_{1.1}$Se$_{0.6}$Te$_{0.4}$ [62], (c) LiFeAs [66], (d) Tl$_{0.58}$Rb$_{0.42}$Fe$_{1.72}$Se$_2$ [72].

As shown in Fig. 7(a), the electron-doped compound Ba(Fe$_{1-x}$Co$_x$)$_2$As$_2$ behaves similarly to the hole-doped (Ba$_{1-x}$K$_x$)Fe$_2$As$_2$ and the upper critical fields for H//c and H⊥c converge at a similar value in the limit of zero temperature [59-61]. The particular curvature of the temperature dependent $\mu_0H_{c2}(T)$ for H//c is likely attributed to its multi-band electronic structure, each band being with different electronic diffusivity [60]. On the other hand, it was also argued that the Pauli paramagnetic limit may become dominant to suppress superconductivity at low temperatures while the magnetic field is applied along the ab-plane[60].

Similar case also applies to the 11-type iron chalcogenides (see Fig. 7(b)). It was shown that Fe$_{1.11}$Te$_{0.6}$Se$_{0.4}$ possesses a large upper critical field of 45T at zero temperature in spite of its relatively low superconducting transition temperature ($T_c \approx$ 14K) [62]. Such an enhancement of $\mu_0H_{c2}(0)$ is likely attributed to the increased disorder effect arising from the excess irons in this compound. Nevertheless, the upper critical field shows remarkably isotropic behavior at low temperature [62], resembling that of 122-type compounds. Later Lei et al studied the upper critical field of $Fe_{1.02}Te_{0.61}Se_{0.39}$ and $Fe_{1.05}Te_{0.89}Se_{0.11}$ using a DC magnetic field up to 35T and confirmed our conclusion [63]. It was found that its zero-temperature upper critical field is much lower than the orbital limit, and therefore, the paramagnetic effect may play a role on pair-breakings for both H//ab and H//c [63].

The 111-type LiFeAs demonstrates simple metallic behavior in the normal state without showing evidence of a structural or magnetic phase transition [11-13]. The upper critical field of LiFeAs has been consistently obtained by means of measuring the magnetic torque [64], resonant frequency shift based on the tunnel diode oscillator [65] and the electrical resistivity [66, 67]. Among the FeSCs, LiFeAs shows a relatively low value of $\mu_0H_{c2}(0)$ which reaches 15T for H//c and 24T for H//ab, allowing us to study $\mu_0H_{c2}(T_c)$ in the full temperature range without any extrapolation. As shown in Fig.7(c), one can see that the upper critical field $\mu_0H_{c2}(T_c)$ for H//ab can be described by the WHH model only after including the spin paramagnetic effect. On the other hand, the WHH model can well describe the experimental data of $\mu_0H_{c2}(T_c)$ for H//c without considering the spin paramagnetic effect, from which $\mu_0H_{c2}^{orb}(0)$ is estimated to be 14.5T following Eq. 5. Furthermore, it was also proposed that a two-band model which takes into account the possibility of the Fulde-Ferrel-Larkin-Ovchinnikov state might be able to illustrate the upper critical field in LiFeAs [65].

Superconductivity was recently discovered in the iron selenides $AFe_xSe_2$ with patterned Fe-vacancy which $T_c$ reaches up to 33K [14-18]. Even though the electronic band structure [68-70] and the characters of their parent compounds [29-32] are considerably different from other FeSCs, $\mu_0H_{c2}(T_c)$ of the iron selenides bears many similarities to that of other families [71,72], indicating universal behavior of the upper critical field in FeSCs. As an example, we show the temperature dependence of $\mu_0H_{c2}(T_c)$ for $Tl_{0.58}Rb_{0.42}Fe_{1.72}Se_2$ in Fig. 7(d) [72]. $\mu_0H_{c2}(T_c)$ linearly increases with decreasing temperature for H//c, reaching $\mu_0 H_{c2}^{H\parallel c}(0K) \simeq 60T$. On the other hand, a larger upper critical field ($\mu_0 H_{c2}^{H\parallel ab}(18K) \approx 60T$) with a strong convex curvature is observed for H//ab. Analysis based on the WHH model indicates that the upper critical field $\mu_0 H_{c2}(0)$ is orbitally limited for H//c, but is likely limited by the spin paramagnetic effect for H//ab.

Table 2: Upper critical fields and some related parameters of several FeSCs

| Compounds | Field Orientation | $T_c$ (K) | $\left(\dfrac{dH_{c2}}{dT}\right)_{T_c}$ (T/K) | $H_{c2}^{orb}(0)$ (T) | $\mu_0 H^P(0)$ (T) | $\mu_0 H_{c2}^{exp}$ (T) | $\xi(0)$ (nm) | $\alpha$ | $\gamma(T_c)$ | $\gamma(T_{min})$ | Ref. |
|---|---|---|---|---|---|---|---|---|---|---|---|
| $Tl_{0.58}Rb_{0.42}Fe_{1.72}Se_2$ | H//c | 33 | 2 | 45 | 60.6 | 52(4K) | - | - | 8.1(32K) | 2.4(20K) | [72] |
| | H//ab | | 12 | 273 | | 54(18K) | - | 5.6 | | | |
| $NdFeAsO_{0.82}F_{0.18}$ | H//c | 47 | 1.3 | 42 | 85.6 | 43(18K) | 0.26 | - | 6(46K) | 5.4(35K) | [48] |
| | H//ab | | 6.5 | 210 | | 57(34K) | 2.3 | 3.5 | | | |
| $SmFeAsO_{0.85}$ | H//c | 50 | 2.5 | 84 | 93.9 | 56(27K) | 3.6 | - | 5(49K) | 3.5(43K) | [55] |
| | H//ab | | 11 | 378 | | 51(43K) | 17 | 2.3 | | | |
| $Ba(Fe_{1-x}Co_x)_2As_2$ (x=0.08) | H//c | 25 | 2 | 34.5 | 46.5 | 40(4.2K) | 1.48 | - | 2.7(22K) | 1.4(6K) | [59,60] |
| | H//ab | | 6 | 103.5 | | 53(6K) | 2.45 | 2.03 | | | |
| LiFeAs | H//c | 18 | 1.2 | 14.5 | 32.6 | 15(0K) | 1.7 | 0 | 2.5(14K) | 1.49(1.4K) | [64-67] |
| | H//ab | | 3.3 | 39.8 | | 24.2(0K) | 4.8 | 1.74 | | | |
| $(Ba, K)Fe_2As_2$ | H//c | 28 | 2.9 | 56 | 52.4 | 55(9K) | 2.17 | 1 | 2(27K) | 1.1(12K) | [49,56] |
| | H//ab | | 5.4 | 104 | | 57(10K) | 2.17 | 1.9-2.2 | | | |
| $Fe_{1.1}Se_{0.6}Te_{0.4}$ | H//c | 14 | 3.8 | 36.9 | 26 | 47 (0K) | 2.65 | 0.88 | 2(12K) | 0.94(0.5K) | [62,63] |
| | H//ab | | 8.9 | 86 | | 47 (0K) | 2.65 | 2.3-3.2 | | | |

For comparison, we summarize the upper critical fields and the related fitting parameters of a few typical FeSCs in Table 2, in which $H_{c2}^{orb}(0)$, $\mu_0 H^P(0)$, $\mu_0 H_{c2}^{exp}$, $\xi$, $\alpha$, $\gamma(T_c)$ and $\gamma(T_{min})$ represent the upper critical fields in the orbital limit, the Pauli limit, the experimental values of the upper critical field, the coherence length, the Maki parameter, the anisotropic parameter near $T_c$ and in the available minimum temperatures, respectively. The upper critical fields, normalized to the corresponding orbital limiting values, $\mu_0 H_{c2}^{orb}(0K)$, are shown in Fig.8. One can see that various FeSCs demonstrate remarkably universal behavior. In the case of H//c, the normalized upper critical field for various families of FeSCs almost collapse on the same curve, showing a linear increase with decreasing temperature. The derived upper critical field approaches or falls slightly below the corresponding orbital limit at zero temperature, indicating that the Cooper pairs are dominantly destroyed by the orbital effect for field applied along the c-axis. Distinct features are observed when the magnetic field is applied in the ab-plane (see Fig. 8(b)). In this field orientation, the upper critical field $\mu_0 H_{c2}(0)$ is suppressed much below the corresponding orbital limits. Nevertheless, the temperature dependence of $\mu_0 H_{c2}^{H\|ab}(T_c)$ can be nicely described in terms of the WHH model after considering the spin paramagnetic effect (Eq. 5). The derived results are shown in Fig. 8(b) by assuming $\lambda_{so}=0$. One can see that the curvature is quite sensitive to the Maki parameter $\alpha$, which value reaches the highest in $Tl_{0.58}Rb_{0.42}Fe_{1.72}Se_2$ and the lowest in (Ba,K)Fe$_2$As$_2$ and LiFeAs. The initial slope of $\mu_0 H_{c2}(T_c)$ near $T_c$ is proportional to $(lv_F)^{-1}$ [67], where $v_F$ is the Fermi velocity and $l$ is the mean free path. Thus the upper critical field may be enhanced by introducing disorder. Indeed, the variance of the upper critical field and the corresponding Maki parameter $\alpha$ as shown in Table 2 and Fig. 8(b) may be argued in terms of the disorder effect. According to the residual-resistivity ratio, LiFeAs [66] is much cleaner than $Fe_{1.11}Te_{0.6}Se_{0.4}$ [62] and $Tl_{0.58}Rb_{0.42}Fe_{1.72}Se_2$ [18]. Correspondingly, the orbital limiting field is enhanced in the latter compounds and the Pauli paramagnetic effect may then become important on suppressing superconductivity. Nevertheless, the multi-band characters of the FeSCs may largely increase the complexity in analyzing the upper critical field behavior [58].

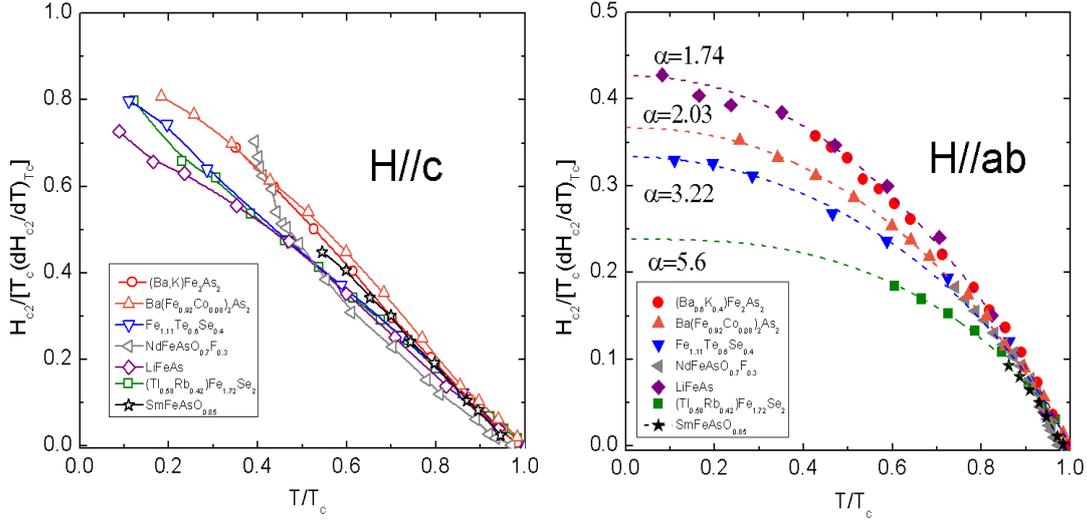

Fig.8: The normalized upper critical field $\mu_0 H_{c2}/[T_c(dH_{c2}/dT)_{T_c}]$ versus the normalized temperature $T/T_c$ for single crystalline FeSCs: (a) H//c, (b) H//ab. The dotted lines in (b) show the fittings based on the WHH model [54-56,59,62,66,72].

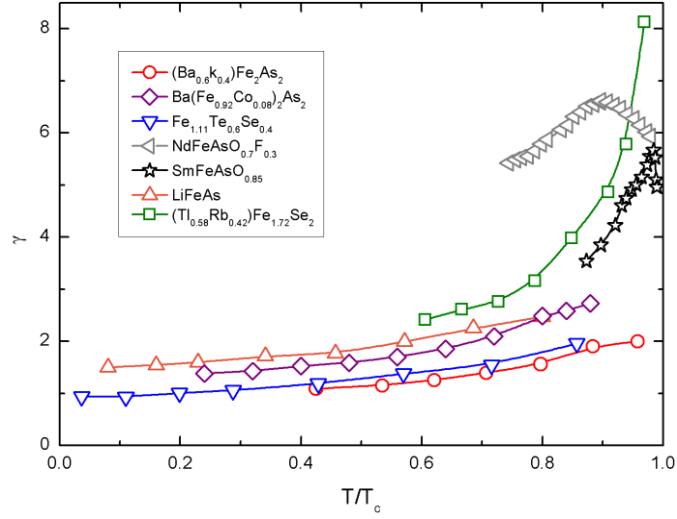

Fig.9: Temperature dependence of the anisotropic parameter γ($T/T_c$) for various FeSCs. These data are from Ref [54-56,59,62,66,72].

**5. Anisotropy of the upper critical field**

In this section, we discuss the anisotropic parameter, defined as $\gamma = H_{c2}^{H\perp c} / H_{c2}^{H\parallel c}$, for some of the FeSCs, which temperature dependence is shown in Fig. 9 and also in Table 2. One can see that, near $T_c$, the anisotropic parameter is moderate (γ=5~8) for the iron selenides $AFe_xSe_2$ and the 1111-families [54,55,72], and small (γ=2~3) for other FeSCs [56,59,62,66]. However, the anisotropic parameter decreases with decreasing temperature in all the FeSCs, approaching γ=1~2 in the low temperature limit. It is noted that a higher magnetic field is still desired in order to look into the anisotropic behavior at low temperatures for some of the FeSCs. Nearly isotropic upper critical field is a particular feature of the FeSCs since one usually expects anisotropic superconductivity in layered superconductors attributed to their anisotropic electronic structure. In FeSCs, moderate anisotropy is indeed observed in the normal state resistivity. For example, the electrical resistivity along the c-axis can be as large as up to 100 times of the in-plane resistivity [38, 73, 74]. The ARPES experiments also revealed a quasi-two-dimensional electronic structure in some of the FeSCs, in particular for the 1111 type of compounds [75]. Observation of nearly isotropic superconductivity in the FeSCs represents an interesting physical phenomenon which underlying mechanism remains unclear.

The nearly isotropic superconductivity in FeSCs is in sharp contrast to other layered superconductors, including the high-$T_c$ cuprates [76] and the organic superconductors [77]. In these compounds, a quasi-2D electronic structure with a cylinder-like Fermi surface is usually observed. In this case, magnetic fields applied exactly within the conducting planes cannot induce significant circulating currents, as the Fermi-surface cross-sections perpendicular to this are not closed. This prevents orbital mechanisms from limiting the upper critical field and instead superconductivity is mainly suppressed by the spin paramagnetic effect, leading to a sharp increase of the upper critical field. A typical example of such a scenario comes from the organic superconductors. As an example, Fig.10 presents the angle dependence of the upper critical field $\mu_0H_{c2}(\theta)$ for the organic superconductor κ-(BEDT-TTF)$_2$Cu(NCS)$_2$ at 1.4K [77], which clearly shows a cusp feature for the in plane magnetic fields ($\theta=\pm 90°$).

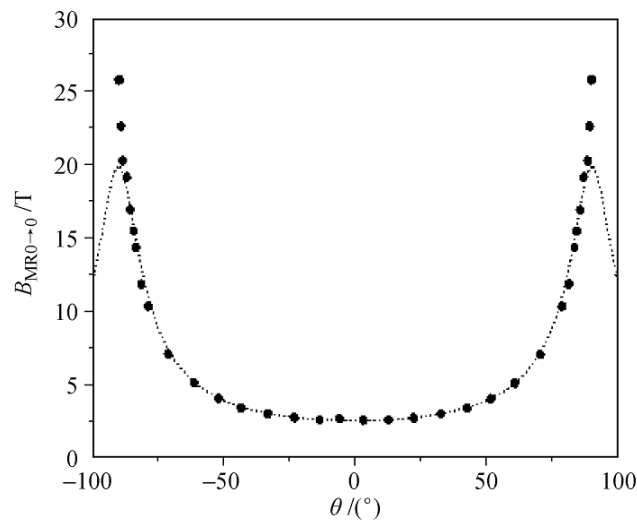

Fig.10 Angle dependence of the upper critical field $\mu_0H_{c2}(\theta)$ at 1.4K for κ-(BEDT-TTF)$_2$Cu(NCS)$_2$ [77]

As mentioned above, the physical origin of the nearly isotropic upper critical field in FeSCs is not yet clear. The following mechanisms might contribute to such unique

properties:

First, the electronic structures of the FeSCs are more three dimensional in comparison with other layered superconductors. The recent ARPES experiments have confirmed that there exists significant dispersion in the $k_z$-direction, providing more direct evidence for a 3D-like superconductors [78,79]. In addition, the coherent length is comparable or even larger than the distance between FeAs layers, which may result in significant inter-layer couplings.

Second, the Pauli paramagnetic effect may play an important role. As we have shown in Section 4, the pair breaking is likely dominated by spin paramagnetic effect for H//ab and by orbital effect for H//c in the iron based superconductors. Usually the orbital pair breaking is more effective near $T_c$, whereas the limiting effect can be caused by Zeeman splitting with increasing magnetic field. At low temperatures, the Pauli paramagnetic effect may become strong enough to compensate the orbital pair-breaking mechanism that makes $\mu_0 H_{c2}^{H\|ab}$ and $\mu_0 H_{c2}^{H\|c}$ get close to each other. Hence, the upper critical field becomes more isotropic at low temperatures [59].

Third, the nature of multi-band superconductivity in FeSCs may complicate the behavior of upper critical field and hence change its anisotropy. As previously discussed in the case of $MgB_2$, the curvature of the upper critical field can be significantly modified depending on the contribution of the various superconducting energy gaps and the concentration of disorder [58]. The upturn curvature for $\mu_0 H_{c2}^{H\|c}(T_c)$ at low temperature might be attributed to the opening of a small gap at low temperature.

## 6. Summary and prospects

In the past three years, intensive efforts have been made to search for new FeSCs with aiming at enhancing the superconducting transition temperature as well as to understand the pairing mechanism of these superconductors. Measurements of the upper critical field using a pulsed magnetic field have shown universal behavior of the upper critical field in all the FeSCs. The findings of nearly isotropic superconductivity in the layered FeSCs are a unique feature among the layered superconductors which might provide insights and restrictions on the theoretical analysis and model of the Fe-based superconductors. Since the FeSCs are multi-band superconductors with hole- and electron-pockets, which configurations may strongly depend on the doping concentrations, it is highly desired to systematically study the doping-dependence of the upper critical field in order to further characterize the superconducting state and the pair-breaking mechanisms.

A large and isotropic upper critical field is required for potential application of a superconductor. The FeSCs exactly meet these requirements and are good candidates for future applications, in particular if its $T_c$ can be further enhanced. Indeed, research efforts on the applied aspects of FeSCs have been launched. For example, superconducting wires have been successfully synthesized [80]. On the other hand, observation of nearly isotropic superconductivity in the FeSCs also suggested that reduced dimensionality in these compounds is not a prerequisite for 'high-temperature' superconductivity, providing an alternative guidance for searching for higher-$T_c$ superconductors.

**Acknowledgement**

The work was supported by the National Science Foundation of China (grant Nos: 10874146, 10934005), the National Basic Research Program of China (973 Program) (2009CB929104, 2011CBA00103), the PCSIRT of the Ministry of Education of China, Zhejiang Provincial Natural Science Foundation of China, and the Fundamental Research Funds for the Central Universities.